# Spatial Control of Frost Formation on Surfaces with Millimetric Serrated Features

Yuehan Yao[1], Emma Feldman[2], Kyoo-Chul Park[3*]

[1]Department of Materials Science and Engineering, Northwestern University, Evanston, IL 60208, USA;
[2]Department of Chemical and Biological Engineering, Northwestern University, Evanston, IL 60208, USA;
[3]Department of Mechanical Engineering, Northwestern University, Evanston, IL 60208, USA.

**Abstract:** Numerous studies have focused on a low surface energy coating and a micro/nanoscale surface texture to design functional surfaces that delay frost formation and reduce ice adhesion. However, the scientific challenges for in developing icephobic surfaces have not been fully addressed because of degradation such as mechanical wearing. Inspired by the suppressed frost formation on concave regions of natural leaves, here we report findings on the frosting process on hydrophobic surfaces with various serrated structures. Dropwise condensation, the first stage of frosting, is enhanced on the peaks and suppressed in the valleys when the serrated surface is exposed to humid air, causing frosting to initiate from the peak. The condensed droplets in the valley are then evaporated due to the different equilibrium vapor pressure of ice and water, resulting in a non-frost band on both hydrophobic and superhydrophilic surfaces. The frost growth is systematically studied by employing various levels of ambient humidity, surface wettability, and surface geometry. Numerical simulations show the critical role of diffusion of water vapor in the formation of the discontinuous frost pattern.

**Introduction**

Ice accretion on surfaces can cause serious energy waste and safety threats in many practical scenarios.[1-7] Air drag increases when ice accumulates on aircrafts or wind turbines and hence disturbs the smooth airflow around them.[8, 9] Heat exchange efficiency can be decreased due to the low thermal conductivity of ice cover.[10] Ice formation can be resulted from freezing of subcooled liquid water on the surfaces or frosting of moisture in the ambient air.[11-13] A lot of research efforts have been devoted to development of anti-icing and de-icing strategies. Supehydrophobic surfaces, which incorporate an extremely low surface energy coating and surface roughness in micro/nanometers, are effective in delaying ice formation because of the low number of nucleation sites and minimized water-solid heat transfer by the air pockets in the space between asperities underneath water.[14-17] However, ice formation is inevitable on such surfaces when the water vapor in the humid air can easily diffuse between the small asperities.[18, 19] This so-called Wenzel ice which eventually forms the ice-solid interface shows a significantly high removal strength due to interlocking between the ice and the surface texture.[20, 21] To minimize the ice removal strength, an additional layer of fluid that is not miscible with water and has a low melting point is often introduced between the ice and solid surface. Slippery liquid-infused porous surfaces (SLIPS), which use a low surface tension oil as the fluid and show molecular level smoothness, are not only able to delay freezing, but also to reduce ice adhesion strength by up to two orders of magnitude compared to the non-coated surface.[22-25] On the other hand, the frozen ice can also be self-lubricated by a thin layer of water between the ice and surface if the surface is coated by a hygroscopic material such as polyethylene oxide brushes.[26-28] Based on these discoveries, other novel surfaces including ferrofluids and oil-infused polymeric materials have also been investigated.[29-31] However, a robust surface that shows long-term resistance to frosting has not been developed.

Frosting is an interfacial process that can be ubiquitously found in nature when the surface temperature drops down to a certain point.[32-34] Figure 1 shows the frost pattern on a fresh American

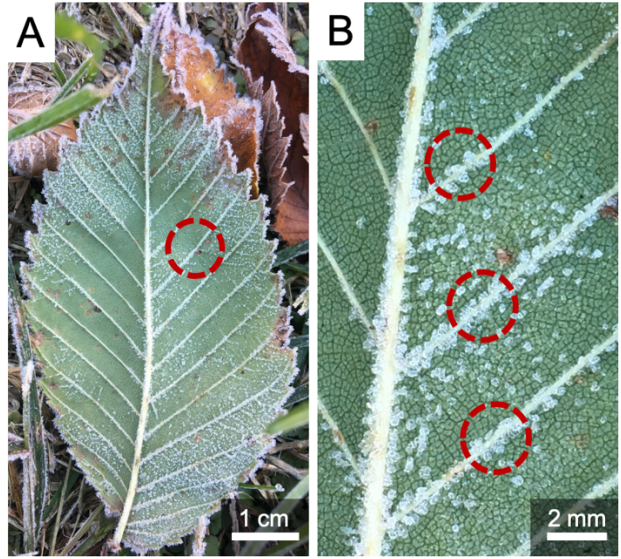

**Figure 1.** (A, B) Non-continuous frost pattern on an American Elm leaf under natural frosting conditions. Red dashed circles indicate the preferential frost formation on the leaf veins.

Elm leaf in a natural environment of frosting. The number density of ice crystals is noticeably high on the veins which are topographically convex, while the flat region between the veins are almost non-frosted. Our previous studies have shown that the millimetric surface topography plays an important role in dropwise condensation, in that droplet growth is enhanced on bumps while suppressed in dimples.[35, 36] The similarity between the convex and concave surface topography formed on natural leaves and them on the bumps and dimples inspired us to investigate the impact of millimetric surface topography on frosting. Here, we study the pattern of frost on non-flat aluminum surfaces with various millimetric serrated features that resemble the convex leaf veins.

**Methods**

**1. Fabrication of aluminum surfaces with serrated patterns.**

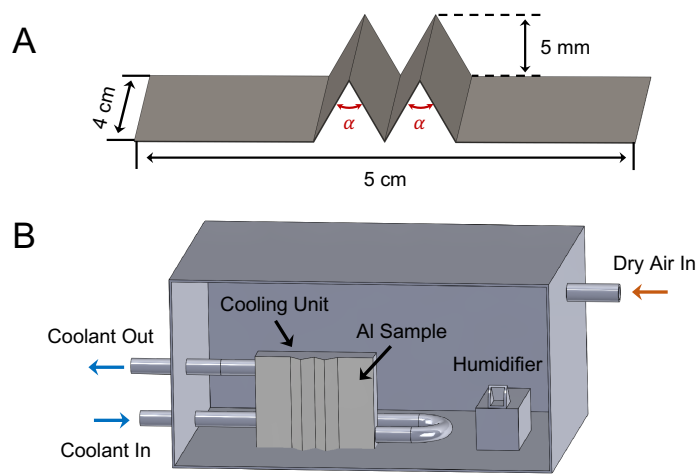

**Figure 2.** Schematics of (A) an aluminum wavy surface with defined geometry, and (B) the experimental setup for controlled frost growth (not drawn in scale). A humidity sensor connected to an external humidity controller is not drawn.

The serrated surfaces with various vertex angles defined in Figure 2A ($\alpha = 40°, 60°, 90°$, and $100°$) were fabricated by a simple molding procedure. The molds with corresponding geometric designs were first printed out by a 3D printer (Form 2, Formlabs, clear resin). The surface patterns were then transferred to a thin aluminum sheet (0.127 mm in thickness, McMaster-Carr) by pressing it between the molds. The height of the peaks is kept to be 5 mm, and the length and width of the samples are 50 and 40 mm, respectively. The geometric parameters are illustrated in Figure 2A.

**2. Fabrication of hydrophobic and superhydrophilic serrated surfaces.**

The patterned aluminum surfaces are intrinsically hydrophilic with a static contact angle of 81°. The patterned samples were then cleaned by oxygen plasma for 1 min to remove the organic contaminants. To obtain hydrophobicity, the cleaned samples were immersed in 1 wt. % solution of the fluoroaliphatic phosphate ester fluorosurfactant (FS-100, Pilot Chemical) in ethanol at 70 °C for 30 min. Superhydrophilic samples were made by boiling the cleaned surfaces in water for 30 min to undergo Boehmitization process. The static contact angles on the superhydrophilic and hydrophobic surfaces are measured to be 0° and 123°, respectively.

**3. Controlled frosting measurement.**

The visualization of the frosting process was done inside a customized chamber (107 cm × 38 cm × 30 cm, L × W × H, Figure 2B). The relative humidity level inside the chamber was maintained by using an electronic humidity controller (Model 5100-240, Electro-Tech Systems Inc.) connected with an ultrasonic humidifier (Pure Enrichment) and an external source of dry air. The air velocity inside the chamber was kept at a low level (< 0.1 m/s) to minimize the effects of convection. Samples were taped (3M$^{TM}$ Scotch Double-Sided Conducted Copper Tape, 12.7 mm wide and 0.04 mm thick) onto a 3D printed cooling unit, one side of which has the same pattern as the aluminum sample. A small wall thickness of 0.5 mm was used to minimize the temperature difference across the aluminum surface. The cooling units were connected to an external circulating liquid chiller (7L AP, VMR) to keep the surface temperature of the samples at -12 ± 0.3°C, which was measured by a digital thermometer (HH66U, OMEGA). Ambient temperature was 23.5 ± 0.5°C. The samples were vertically positioned, and a plastic cover was used to isolate the surfaces from the ambient air before the temperature of the sample surface was stabilized, at which $t = 0$ is defined. The frosting processes were recorded using a Nikon D5500 camera with a macro lens (Nikon AF-S DX Micro NIKKOR 40 mm f/2.8G) and a handheld digital microscope (Dino-Lite Premier AF3113T). Figure 2B shows a schematic of the experimental setup for the controlled frosting experiments.

**4. Numerical simulation of diffusion by using COMSOL Multiphysics©.**

Models for steady state transport of dilute species were used to numerically simulate the diffusion of water vapor near the serrated features. 2-D coordinates were employed for all simulations by neglecting the effect of finite length of the samples in the direction of protrusion. The models were built using the cross-sectional geometries of the serrated surfaces to be studied. Boundary conditions were chosen depending on the specific stage to be simulated. See Results and Discussions for the boundary conditions used for each simulation.

**Results and Discussions**

**1. Discontinuous frost growth.** Figure 3 shows the time-lapse images of frosting process on a hydrophobic serrated surface with a vertex angle of $\alpha = 60°$, which is defined in Figure 2B, at an ambient humidity of $RH = 25\pm2\%$. Four stages can be clearly identified: I) haze features grow on the reflective surface ($t = 80$ sec), II) frost initiates from the peaks and then quickly propagates towards valley ($t = 200$ sec), III) frost propagation slows down and haze in the valley diminishes ($t = 910$ sec), and IV) ice crystals grow slowly into a dendrite shape ($t = 2270$ sec and onward). With repeated experiments, two traits can be identified which are distinct from the frosting process on a flat surface. At the onset stage ($t < 80$ sec), frosting always initiates from the peaks. At the end stage ($t > 2270$ sec), frost preferentially covers the peaks, while a non-frost region which spans about half of the total surface area between the two peaks exists at the valley. This discontinuous coverage of frost resembles the frost pattern found on the natural leaves, where frost preferentially covers the leaf veins while the flat regions in between show less frost coverage. Since the rate of frost invasion into the valley becomes significantly slower than that of the frost propagation from the peaks, this non-frosted band in the valley can be considered to resist against frost formation for a long period of time.

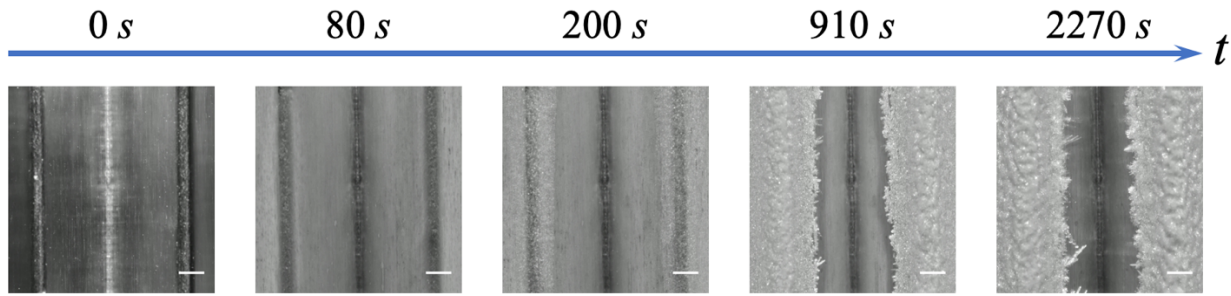

**Figure 3.** Time lapse images showing the condensation (80 s), fast propagation (200 s), evaporation (910 s), and deposition (2270 s) stages. Surface is hydrophobic, and has a vertex angle of 60°. Ambient humidity is 25% at 23.5°C. Scale bars are 1 mm.

The multistage frosting process was verified by visualization under a microscope objective lens as shown in Figures 4. It should be noted that the sample under the microscope lens has a vertex angle $\alpha = 90°$ for better imaging, but it was tested under the same conditions as in Figure 3. Due to the supersaturated conditions in the ambient, dropwise condensation occurs across the whole surface at the first place, as shown in Figures 4A-1 and 4B-1. However, the size of supercooled drops is found to decrease from the peak to the valley, ranging from around 40 $\mu$m near the peak to less than 10 $\mu$m near the valley (in yellow dashed circles). Such distribution of droplet sizes agrees with our previous observation of condensed droplets are larger on bumps than on dimples.[35, 36] These condensed droplets in micrometers effectively diffuse light and explain the "hazy features" found in Figure 3. Following the first condensation stage, frosting initiates from the peak and quickly propagates without changing the pattern formed by the droplets (Fig 4A-2). This observation implies that the frost propagates by interconnecting adjacent droplets, which agrees with the "ice-bridging" mechanism described in other studies.[16, 37, 38] When the ice front approaches the valley, droplets start to varnish. A gap forms between the ice front and the droplets near the valley as shown in Figure 4B-2. This gap prevents the invasion of frost further down to the valley. All the droplets on the right of the gap in Figure 4B-2 eventually disappear in Figure 4B-3, corresponding well with the disappearance of haze shown in Figure 3 ($t = 910$ sec to $t =$ 2270 sec). With the depletion of droplets in the valley, the frost front continues to slowly grow and becomes dendritic, which indicates the droplets may evaporate and then deposit onto the ice. This process is driven by the local concentration gradient of water vapor established by the different vapor pressure of ice and supercooled water. The multi-stage mechanism of frosting is schematically demonstrated in Figure 4C.

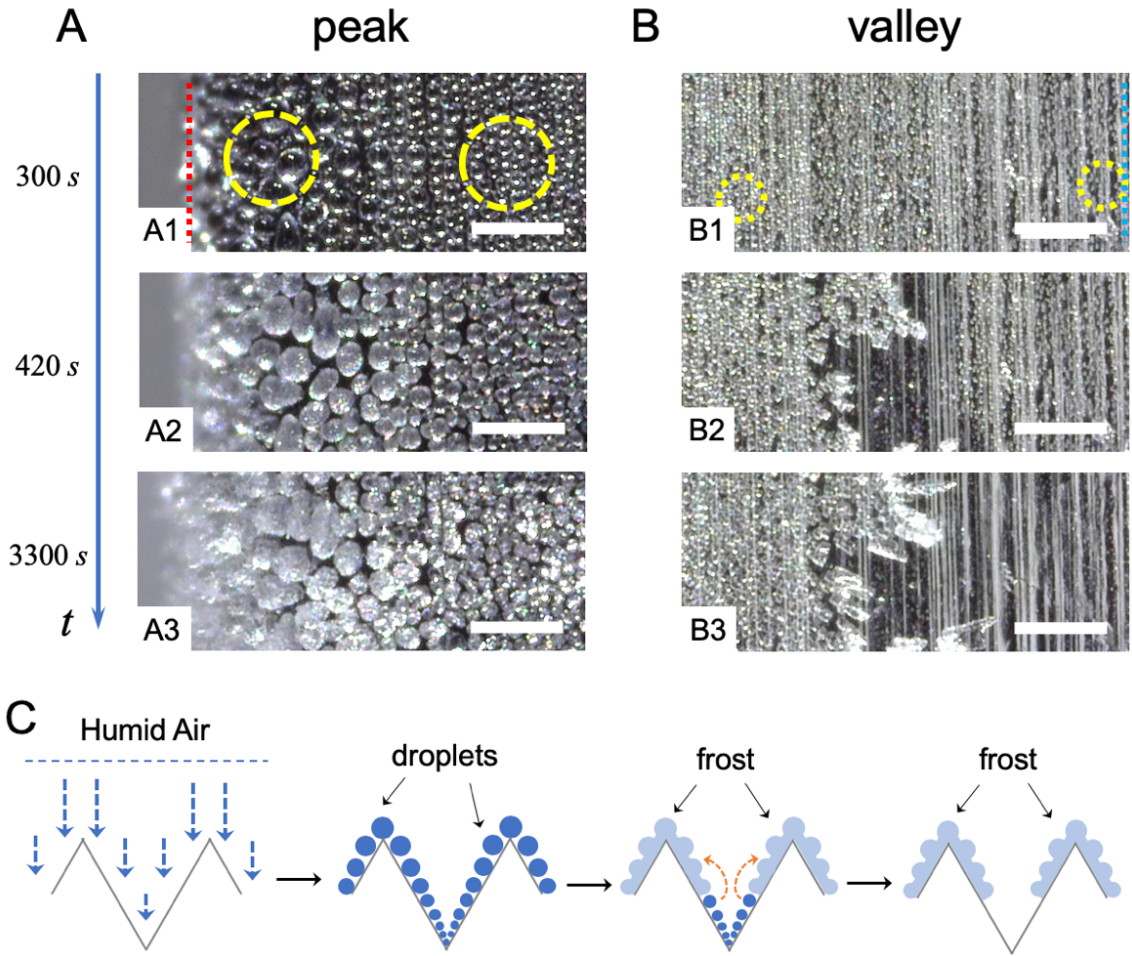

**Figure 4.** Micro images of the frosting process at the (A) peak and (B) valley. Surface has a vertex angle of 90°. Red and blue dotted lines indicate the relative positions of the peak and valley, respectively. Surface is hydrophobic, and subject to $RH = 25\%$ at 23.5°C. Scale bars are 0.2 mm. (C) Schematics of the frosting mechanism.

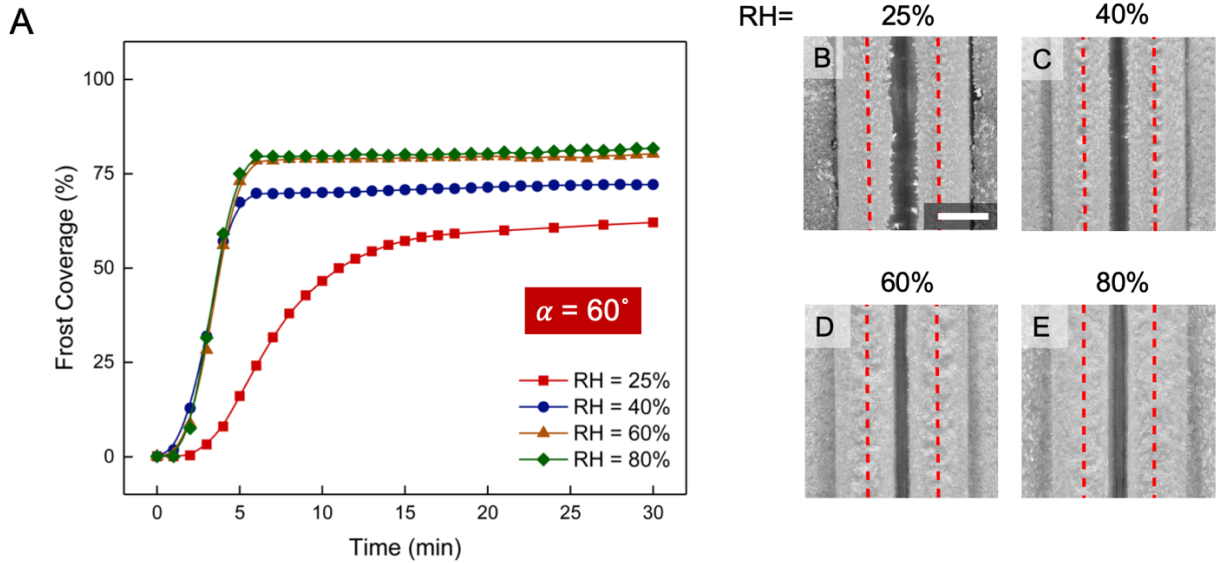

**Figure 5.** (A) Time evolution of frost coverage on a hydrophobic serrated surface with $\alpha = 60°$ under various ambient humidity levels at 23.5°C. Frost pattern at $t = 30$ min for surfaces subject to RH = (B) 25%, (C) 40%, (D) 60%, and (E) 80%. Red dashed lines indicate peaks. Scale bars represent 4 mm.

With the surface geometry controlled to be $\alpha = 60°$, samples were tested at four levels of ambient relative humidity $RH = 25\%$, 40%, 60% and 80%. Figure 5A shows the frost coverage as time progresses. The non-frosted area shrinks as the ambient humidity levels up as shown in Figure 5B-E. The propagation and evaporation stages are noticeably slower for $RH = 25\%$ than the other higher $RH$ levels.

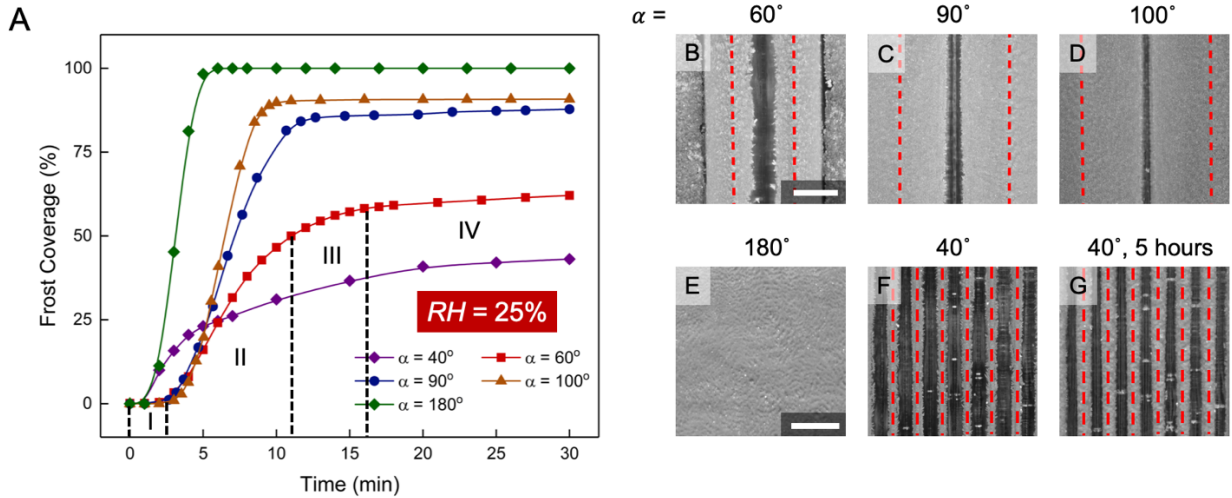

**Figure 6.** (A) Time evolution of frost coverage on hydrophobic serrated surface with various vertex angles. Ambient humidity is 25% at 23.5°C. (B-F) Frost patterns at $t$ = 30 min show the non-frosted band width decreases as $\alpha$ increases. (G) 49% of the total surface area is non-frosted on the surface with $\alpha$ = 40° after 5 hours of experiments. Red dashed lines indicate peaks. The vertex angles ($\alpha$) are shown in insets. Scale bars are 4 mm in (B, C, D), and 8 mm in (E, F, G).

To quantitatively study the impact of surface geometry on the frost pattern, surfaces with various vertex angles were tested under $RH$ = 25%. For samples with $\alpha$ = 60°, 90°, and 100°, the frost coverage was evaluated between the two peaks since the surfaces were designed to have only two peaks. For $\alpha$ = 40° and 180° (i.e., flat surface), the whole surface area is considered. The time evolution of frost coverage is plotted in Figure 6A, and the frost patterns at $t$ = 30 min are shown in Figures 6B-F. The flat surface ($\alpha$ = 180°) shows the lowest resistance against frosting as it was completely covered by frost after 5 min of experiments. The surface with a smaller vertex angle shows better anti-frosting performance as indicated by the lower frost coverage stabilized at $t$ = 30 min. The surface with the smallest vertex angle ($\alpha$ = 40°) shows a frost coverage less than 45% after 30 min. After 5 hours of experiments, the non-frosted area is still 49% (Figure 6G). The four stages described previously can also be identified in each curve with different time spans. The division of stage I-IV shown in Figure 6A is based on $\alpha = 60°$ (red square). All the surfaces start with a relatively short incubation period without any frost coverage, which corresponds to the condensation stage (Stage I). Then a stage where the frost coverage almost soars with time follows, indicating frost quickly propagates (Stage II). The transition period (Stage III) where the slope of the curves in Figure 6A decreases from the linear portion in Stage II to almost 0 indicates the droplets in the valley evaporate and accumulates onto the ice front. Finally, at Stage IV, frost mainly grows out of plane which increases the thickness of the frost cover, while the in-plane frost coverage shows a plateau regime suggesting that the depletion of liquid drops in the valley.

Different from the dropwise condensation on hydrophobic surfaces, water condensate spreads out as a thin film on superhydrophilic surfaces where the $\theta^* \approx 0°$. Figure 7 shows the frost coverage between two peaks on a superhydrophilic surface with a vertex angle of $\alpha = 60°$, and the ambient humidity was kept to be $RH = 25\%$. Similar to the frost pattern observed on the hydrophobic surface, the valley is non-frosted. However, the propagation and evaporation stages (Stage II and IV) are absent. Instead, the condensed liquid quickly freezes ($t < 1$ min). This is presumably because the superhydrophilic surfaces facilitates nucleation of water by providing a high number density of high energy sites.[39, 40] The interdroplet distance is thereby shorter, and nucleated droplets quickly spread out due to the low $\theta^*$, interconnect, and become filmwise. The evaporation stage also becomes transient given both the suppressed condensation in the valley proved by the small droplet sizes shown in Figure 4C-1, and the lack of time for droplet growth caused by the fast freezing. The slow increase in frost coverage (70% at $t = 0.5$ min and 73% at $t = 30$ min) agrees with slopes close to zero at Stage IV shown in Figures 5A and 6A. However, the frost coverage on the superhydrophilic surface at $t = 30$ min (73%, Figure 7B) is greater compared to that on the hydrophobic surface (61%, Figures 5A, 5B) with the same vertex angle and under the same ambient humidity. This can also be explained by the proximity between nucleated drops on superhydrophilic surfaces, which later easily combine to form a liquid film, and freeze collectively.

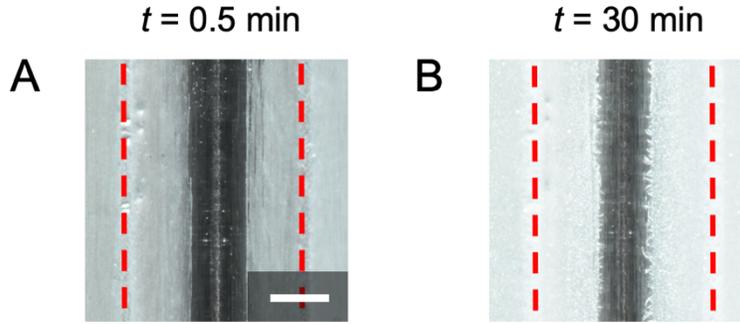

**Figure 7.** Frost pattern on a superhydrophilic surface at $t =$ (A) 0.5 min and (B) 30 min. $\alpha = 60°$. Ambient humidity is 25% at 23.5°C. Red dashed lines indicate peaks. Scale bar is 2mm.

**2. Numerical simulation of mass transfer near the surface features.** Since the surface temperature of the aluminum samples was kept below the dew point under the aforementioned testing conditions, condensation occurs in the first place rather than direct frost deposition by ablimation even though surface temperature is below 0°C ($T_{surf} = -12°C$, $T_{dew} = 1.9°C$ for $RH = 25\%$ and $T_{ambient} = 23.5°C$). Condensation therefore has a huge impact on the following frosting process.

To explain the drop size distribution observed in Figures 4A-B, we numerically simulated the diffusional transport of water vapor near the serrated features. Figure 8A shows the boundary conditions employed. The thickness of the diffusion layer is assumed to be 1 mm according to many previous studies.[41, 42] Concentration near the surface is chosen to be that of the supercooled water at the surface temperature, and concentration at the diffusion boundary is chosen to be that corresponding to the ambient humidity.[43-45] The concentration field of water vapor near the serrated features is shown in Figure 8B. Even though the concentration of water vapor is constant across the surface as defined in the boundary conditions, the concentration isolines are more densely distributed near the peaks while they are sparser near the valley. The denser isoline distribution implies a greater magnitude of concentration gradient $|\nabla c|$ and therefore a greater diffusion flux $J_C$ as captured in Figure 8C. Consequently, the number of molecules to impact onto the peaks is much greater than that into the valley per unit area and time. This explains the larger droplet size and number density at the peaks compared with them at the valleys.

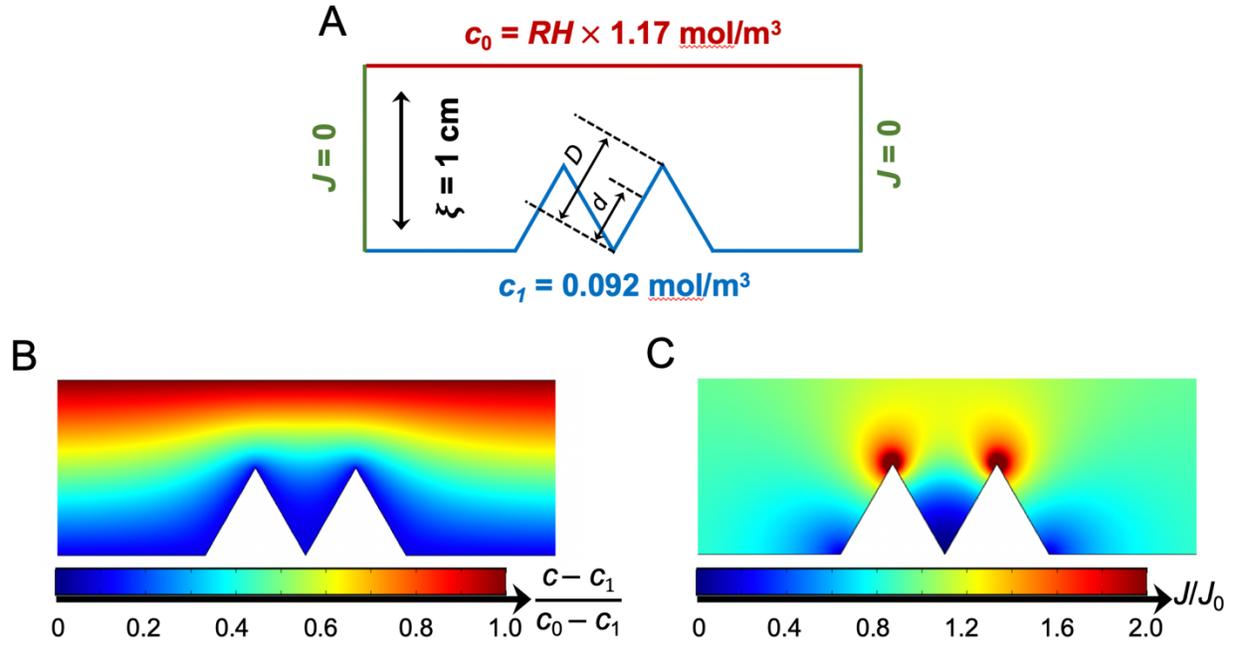

**Figure 8.** (A) Boundary conditions for simulating diffusion of water vapor near the serrated surface during the condensation stage. Distribution of (B) water vapor concentration and (C) flux magnitude near the serrated surface features. The flux filed is normalized by $J_0 = (c_0 - c_1)/\xi$. Drawings are not in scale.

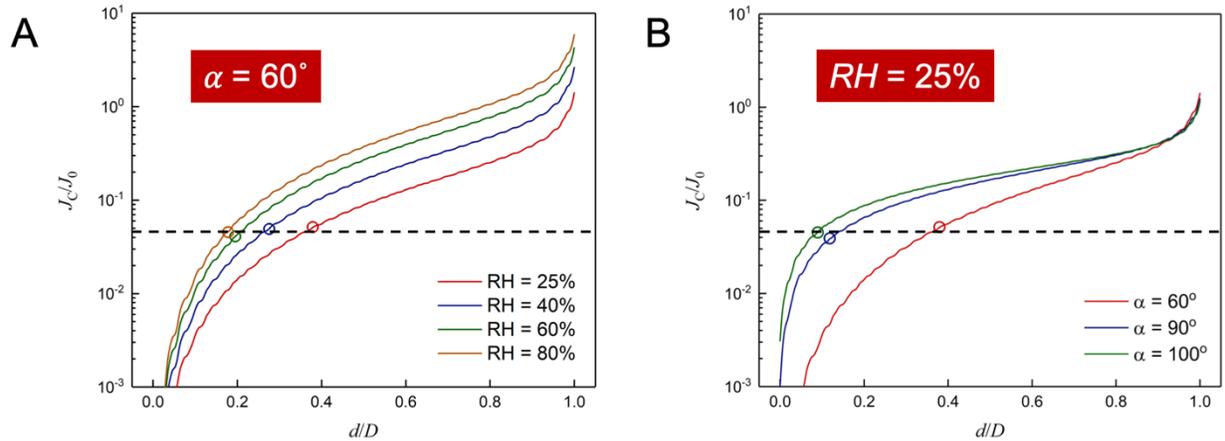

**Figure 9.** Distribution of the normalized diffusion flux from the valley to the peak for (A) ambient $RH$ = 25%, 40%, 60%, and 80% with fixed $\alpha$ = 60°, and (B) $\alpha$ = 60°, 90°, and 100° with fixed ambient $RH$ = 25%. Colored circles indicate similar $J_C/J_0$ at the mean positions of the frost front at $t$ = 30 min (Stage IV) for all levels of $RH$ and $\alpha$.

Following the condensation stage, ice can nucleate in a micro-droplet nearby a larger droplet to initiate propagation of ice front.[37, 38] The likelihood for such micro-sized ice seeds to contact with a large droplet is much higher at the peak due to a higher number density of drops and greater size of them on the peak. Once frosting is initiated, the condensed drops near the ice front evaporate because of the lower vapor pressure of ice. The water vapor then deposits onto the ice front and fills the gap in between until the ice seeds reach nearby larger drops and initiate the freezing of those droplets. As a result, a larger number density of droplets means a shorter distance between the ice front and the nearby drops, and therefore a less volume of the interdroplet space to be filled by the evaporation-deposition mechanism. Hence, a critical droplet size exists below which droplets will varnish before the space between ice and droplets is filled and frost propagation by interconnection will not occur. The low number density and small size of droplets at the valley caused by the extremely small magnitude of diffusion flux captured in Figure 8C result in the complete evaporation of droplets before the ice-bridges reach the droplets, and hence the formation of the non-frosted band in the valley.

The final frost pattern can be affected by the size and number density of the condensed droplets. Figure 9A shows the $J_C/J_0$ increases when the ambient humidity level increases from 25% to 80% for the same type of surface geometry. On the other hand, the location that shows the same $J_C/J_0$ is closer to the valley when the relative humidity rises. It is also possible to estimate the $J_C/J_0$ near the frost front at the deposition stage (Stage IV) by assuming a straight frost front such that $d/D$ of points on that line are identical. The spatial parameter $d/D$ representing the mean frost front, $\tilde{d}/D$ is hence evaluated by $1 - f$, where $f$ is the frost coverage stabilized at $t = 30$ min. The open circles in Figure 9A indicate such $\tilde{d}/D$ at certain $RH$ levels. The ordinate $J_C/J_0$ evaluated at $\tilde{d}/D$, $\widetilde{J_C}/J_0$, is similar for all $RH$ levels (black dashed line, $\widetilde{J_C}/J_0 \approx 3.5 \times 10^{-2}$), which suggests the rate of condensation is similar where the ice front stops regardless of the ambient humidity. Such critical value is also found when the ambient humidity is fixed and the surface geometry is altered, as shown in Figure 9B. It is worth noting that even though $J_C/J_0$ at the peak ($d/D \rightarrow 1$) decreases as the vertex angle increases, it also decreases quickly when the location $d/D$ moves away from the peak. For the vast majority of the flat area between the peak and valley where $d/D < 0.9$, smaller vertex angle shows smaller $J_C/J_0$. It implies that condensation and frosting on most surface area are more suppressed when a smaller vertex angle is used.

**Conclusion**

In summary, we studied the condensation frosting process on hydrophobic surfaces with millimetric serrated features inspired by the veins of natural leaves. Our results show that frosting always initiates from the peak and undergoes a four-stage process including condensation, fast propagation, evaporation, and out-of-plane growth. A discontinuous frost pattern-a non-frosted area centered at the valley-forms and is found to be able to resist further frosting for a long period of time. The spatial span of the non-frosted area is found to expand when the vertex angle of the serrated feature decreases, and when the ambient relative humidity decreases. By simulating the mass transport of water vapor near the serrated features, the distribution of droplet size and number density of drops is explained by the greater magnitude of diffusion flux near the peak than in the valley. We rationale that condensation plays a key role in the formation of the discontinuous frost pattern by comparing the distribution of diffusion flux magnitude and the corresponding frost patterns for the surfaces with various vertex angles and ambient humidity. It is found the frost front for all surface geometries and humidity shares a similar magnitude of diffusion flux, suggesting the frost pattern affected by the millimetric topography originates from the diffusional behavior of water vapor. We envision that this macroscopic surface topography effect can be applied to realize spatial control of frost in a wide spectrum of applications such as aviation, wind power generation, and infrastructure of buildings.

**Reference:**

[1] McKay, G. A.; Thompson, H. A. Estimating the hazard of ice accretion in Canada from climatological data. *J. Appl. Meteorol.* **1969**, *8*, 927-935.

[2] Bernstein, B.C.; Ratvasky, T.P.; Miller, D.R.; McDonough, F. Freezing rain as an in-flight icing hazard. **2000**.

[3] Politovich, M. K. Aircraft icing caused by large supercooled droplets. *J. Appl. Meteorol.* **1989**, *28*, 856-868.

[4] Bonelli, P.; Lacavalla, M.; Marcacci, P.; Mariani, G.; Stella, G. Wet snow hazard for power lines: a forecast and alert system applied in Italy. *Nat. Hazards Earth Syst. Sci.* **2011**, *11*, 2419-2431.

[5] Cole, J.; Sand, W. Statistical study of aircraft icing accidents. *29th Aerospace Sciences Meeting*, **1991**, 558.

[6] Bragg, M. B.; Broeren, A. P.; Blumenthal, L. A. Iced-airfoil aerodynamics. *Prog. Aero. Sci.* **2009**, *41*, 323-362.

[7] Huang, L.; Liu, Z.; Liu, Y.; Gou, Y.; Wang, J. Experimental study on frost release on fin-and-tube heat exchangers by use of a novel anti-frosting paint. *Exp. Therm. and Fluid Sci.* **2009**, *33*, 1049-1054.

[8] Lynch, F. T.; Khodadoust, A. Effects of ice accretions on aircraft aerodynamics. *Prog. Aero. Sci.* **2001**, *37*, 669-767.

[9] Cao, Y.; Ma, C.; Zhang, Q.; Sheridan, J. Numerical simulation of ice accretions on an aircraft wing. *Aero. Sci. Technol.* **2012**, *23*, 296-304.

[10] Yu, J.; Zhang, H.; You, S. Heat transfer analysis and experimental verification of casted heat exchanger in non-icing and icing conditions in winter. *Renew. Energ.* **2012**, *41*, 39-43.

[11] Zhao, Y.; Wang, R.; Yang, C. Interdroplet freezing wave propagation of condensation frosting on micropillar patterned superhydrophobic surfaces of varying pitches. *Int. J. Heat Mass Transf.* **2017**, *108*, 1048-1056.

[12] Varanasi, K. K.; Deng, T.; Smith, J. D.; Hsu, M.; Bhate, N. Frost formation and ice adhesion on superhydrophobic surfaces. *Appl. Phys. Lett.* **2010**, *97*, 234102.

[13] Meuler, A. J.; Smith, J. D.; Varanasi, K. K.; Mabry, J. M.; McKinley, G. H.; Cohen, R. E. Relationships between water wettability and ice adhesion. *ACS Appl. Mater. Interfaces* **2012**, *2*, 3100-3110.

[14] Cao, L.; Jones, A. K.; Sikka, V. K.; Wu, J.; Gao, D. Anti-icing superhydrophobic coatings. *Langmuir* **2009**, *25*, 12444-12448.

[15] Wang, Y.; Xue, J.; Wang, Q.; Chen, Q.; Ding, J. Verification of icephobic/anti-icing properties of a superhydrophobic surface. *ACS Appl. Mater. Interfaces* **2013**, *5*, 3370-3381.

[16] Boreyko, J. B.; Collier, C. P. Delayed frost growth on jumping-drop superhydrophobic surfaces. *ACS Nano* **2013**, *7*, 1618-1627.

[17] Mishchenko, L.; Hatton, B.; Bahadur, V.; Taylor, J. A.; Krupenkin, T.; Aizenberg, J. Design of ice-free nanostructured surfaces based on repulsion of impacting water droplets. *ACS Nano* **2010**, *4*, 7699-7707.

[18] Narhe, R. D.; Beysens, D. A. Nucleation and growth on a superhydrophobic grooved surface. *Phys. Rev. Lett*. **2004**, *93*, 076103.

[19] Wier, K. A.; McCarthy, T. J. Condensation on ultrahydrophobic surfaces and its effect on droplet mobility: ultrahydrophobic surfaces are not always water repellant. *Langmuir* **2006**, *22*, 2433-2436.

[20] Stone, H. A. Ice-phobic surfaces that are wet. *ACS Nano* **2012**, *6*, 6536-6540.

[21] Bengaluru Subramanyam, S.; Kondrashov, V.; Rühe, J.; Varanasi, K. K. Low ice adhesion on nano-textured superhydrophobic surfaces under supersaturated conditions. *ACS Appl. Mater. Interfaces* **2016**, *8*, 12583-12587.

[22] Kim, P.; Wong, T. S.; Alvarenga, J.; Kreder, M. J.; Adorno-Martinez, W. E.; Aizenberg, J. Liquid-infused nanostructured surfaces with extreme anti-ice and anti-frost performance. *ACS Nano* **2012**, *6*, 6569-6577.

[23] Wong, T. S.; Kang, S. H.; Tang, S. K., Smythe; E. J., Hatton, B. D.; Grinthal, A.; Aizenberg, J. Bioinspired self-repairing slippery surfaces with pressure-stable omniphobicity. *Nature* **2011**, *477*, 443.

[24] Stamatopoulos, C.; Hemrle, J.; Wang, D.; Poulikakos, D. Exceptional anti-icing performance of self-impregnating slippery surfaces. *ACS Appl. Mater. Interfaces* **2017**, *9*, 10233-10242.

[25] Chen, J.; Dou, R.; Cui, D.; Zhang, Q.; Zhang, Y.; Xu, F.; Zhou, X.; Wang, J.; Song, Y.; Jiang, L. Robust prototypical anti-icing coatings with a self-lubricating liquid water layer between ice and substrate. *ACS Appl. Mater. Interfaces* **2013**, *5*, 4026-4030.

[26] Daniel, D.; Timonen, J. V.; Li, R.; Velling, S. J.; Aizenberg, J. Oleoplaning droplets on lubricated surfaces. *Nat. Phys.* **2017**, *13*, 1020.

[27] Chen, D.; Gelenter, M. D.; Hong, M.; Cohen, R. E.; McKinley, G. H. Icephobic surfaces induced by interfacial nonfrozen water. *ACS Appl. Mater. Interfaces* **2017**, *9*, 4202-4214.

[28] Li, C.; Li, X.; Tao, C.; Ren, L.; Zhao, Y.; Bai, S.; Yuan, X. Amphiphilic antifogging/anti-icing coatings containing POSS-PDMAEMA-b-PSBMA. *ACS Appl. Mater. Interfaces* **2017**, *9*, 22959-22969.

[29] Irajizad, P.; Hasnain, M.; Farokhnia, N.; Sajadi, S. M.; Ghasemi, H. Magnetic slippery extreme icephobic surfaces. *Nat. Commun.* **2016**, *7*, 13395.

[30] Golovin, K.; Kobaku, S. P.; Lee, D. H.; DiLoreto, E. T.; Mabry, J. M.; Tuteja, A. Designing durable icephobic surfaces. *Sci. Adv.* **2016**, *2*, e1501496.

[31] Yeong, Y. H.; Wang, C.; Wynne, K. J.; Gupta, M. C. Oil-infused superhydrophobic silicone material for low ice adhesion with long-term infusion stability. *ACS Appl. Mater. Interfaces* **2016**, *8*, 32050-32059.

[32] Nath, S.; Ahmadi, S. F.; Boreyko, J. B. A review of condensation frosting. *Nanosc. Microsc. Therm.* **2017**, *21*, 81-101.

[33] Boreyko, J. B.; Hansen, R. R.; Murphy, K. R.; Nath, S.; Retterer, S. T.; Collier, C. P. Controlling condensation and frost growth with chemical micropatterns. *Sci. Rep.* **2016**, *6*, 19131.

[34] Ahmadi, S. F.; Nath, S.; Iliff, G. J.; Srijanto, B. R.; Collier, C. P.; Yue, P.; Boreyko, J. B. Passive Antifrosting Surfaces Using Microscopic Ice Patterns. *ACS Appl. Mater. Interfaces* **2018**, *10*, 32874-32884.

[35] Yao, Y.; Aizenberg, J.; Park, K. C. Dropwise condensation on hydrophobic bumps and dimples. *Appl. Phys. Lett.* **2018**, *112*, 151605.

[36] Park, K. C.; Kim, P.; Grinthal, A.; He, N.; Fox, D.; Weaver, J. C.; Aizenberg, J. Condensation on slippery asymmetric bumps. *Nature* **2016**, *531*, 78.

[37] Jung, S.; Tiwari, M. K.; Poulikakos, D. Frost halos from supercooled water droplets. *Proc. Natl. Acad. Sci. U.S.A.* **2012**, *109*, 16073-16078.

[38] Nath, S.; Boreyko, J. B. On localized vapor pressure gradients governing condensation and frost phenomena. *Langmuir* **2016**, *32*, 8350-8365.

[39] Sheng, Q.; Sun, J.; Wang, Q.; Wang, W.; Wang, H. S. On the onset of surface condensation: formation and transition mechanisms of condensation mode. *Sci. Rep.* **2016**, *6*, 30764.

[40] Merikanto, J.; Vehkamäki, H.; Zapadinsky, E. Monte Carlo simulations of critical cluster sizes and nucleation rates of water. *J. Chem. Phys.* **2004**, *121*, 914-924.

[41] Zhao, Y.; Preston, D. J.; Lu, Z.; Zhang, L.; Queeney, J.; Wang, E. N. Effects of millimetric geometric features on dropwise condensation under different vapor conditions. *Int. J. Heat Mass Transf.* **2018**, *119*, 931-938.

[42] Medici, M. G.; Mongruel, A.; Royon, L.; Beysens, D. Edge effects on water droplet condensation. *Phys. Rev. E* **2014**, *90*, 062403.

[43] Beltramino, G.; Rosso, L.; Smorgon, D.; Fernicola, V. Vapor pressure measurements over supercooled water in the temperature range from $-10^1$ °C to$+ 10^{-2}$ °C. *J. Chem. Thermodyn.* **2017**, *105*, 159-164.

[44] Lide, D. R. *CRC Handbook of Chemistry and Physics*.; Boca Raton: CRC, 2012.

[45] Engineering ToolBox. Air - Diffusion Coefficients of Gases in Excess of Air. https://www.engineeringtoolbox.com/air-diffusion-coefficient-gas-mixture-temperature-d_2010.html [Accessed 02 01. 2019].